\documentclass[12pt]{article}
\usepackage{times}
\usepackage[dvips]{graphicx}
\title{Variable range random walk}

\author{
Takashi Odagaki$^\ast$\\
\\
\normalsize{Kyushu University, Nishiku, Fukuoka 819-0395, Japan}\\
\normalsize{and}\\
\normalsize{Research Institute for Science Education, Inc.}\\
\normalsize{Kitaku, Kyoto 603-8346, Japan}\\
\normalsize{$^\ast$Corresondence to: t.odagaki@kb4.so-net.ne.jp}\\
}

\date{\today}

\begin{document} 

\baselineskip24pt

\maketitle 

%\preprint{RISE/21-02}

\begin{abstract}
Exploiting the coherent medium approximation, random walk among
sites distributed randomly in space is investigated when the jump rate
depends on the distance between two adjacent sites. 
In one dimension, it is shown that when the jump rate decays exponentially
in the long distance limit, a non-diffusive to diffusive transition occurs
as the density of sites is increased.
In three dimensions, the transition exists when the jump rate has a super
Gaussian decay.
\end{abstract}

%\pacs{}
%\keywords{random walk. coherent medium approximation, diffusivity transition}
%Use showkeys class option if keyword

\section{\label{sec:1}Introduction}
Random walk is highly robust which has been applied in almost all areas of science including
biology, materials sciences, economics and social phenomena \cite{mike}.
It has also been exploited in the description of structural evolution by
the free energy landscape theory of non-equilibrium systems \cite{oda-FEL}.
Random walk is generally formulated on lattices and complex networks \cite{oshima-oda}.
In some applications, a random walker is assumed to make jumps among sites
randomly distributed in space \cite{S-L}.
In Mott's variable range hopping model\cite{mott}, carriers are allowed to make 
a long range hopping in disordered media. Diffusion processes have also been
investigated when the diffusivity depends on position and time \cite{metzler}.

In this paper, I investigate a random walk among randomly distributed sites, where
a walker makes a jump from a site to its adjacent neighbor on a limited solid angle
along a fixed number of directions.
The jump rate is assumed to depend on the distance between two sites.
Since sites are distributed randomly, the walker makes jumps with variable ranges.
I obtain the diffusion constant within the coherent medium approximation \cite{oda-lax}
and discuss the possibility of a transition from non-diffusive to diffusive states.

In \S 2, I explain the model in some detail and the method of analysis based
on the coherent medium approximation.
In \S 3, I study the random walk in one dimension for three different types of the jump rate and
discuss non-diffusive to diffusive transition for an extended percolation model.
The random walk in three dimensions is investigated in \S 4 and results are discussed in \S 5,

%%Sect 2
\section{\label{Sec:2}Model and the method of analysis}
I consider a random walk, where the transition probability of the random walker
obeys
\begin{equation}
\frac{\partial P({\bf r}_n,t|{\bf r}_0,0)}{\partial t} = -\sum_m w(|{\bf r}_m-{\bf r}_n|)P({\bf r}_n,t|{\bf r}_0,0) + 
\sum_m w(|{\bf r}_n-{\bf r}_m|)P({\bf r}_m,t|{\bf r}_0,0).
\label{eq:1}
\end{equation}
Here,  $P({\bf r}_n,t|{\bf r}_0,0)$ denotes the transition probability that a random walker
is at the site ${\bf r}_n$ at time $t$ when it started ${\bf r}_0$ at time $t=0$, and 
$w(|{\bf r}_m-{\bf r}_n|)$ is the jump rate of a random walker from site ${\bf r}_n$ to site ${\bf r}_m$.
I assume that $w(|{\bf r}_m-{\bf r}_n|)$ is a function of the distance between ${\bf r}_n$ and ${\bf r}_m$.
Usually, $w(r)$ is assumed to be nonzero within a certain distance. For example, the percolation process on
lattices is modeled by jumps of a random walker within nearest neighbor sites.
In this paper, I introduce an extended percolation model in which a random walker can make a longer jump
with smaller rate beyond the limited distance used for the standard percolation model.

The diffusion constant $D$ is given by
\begin{equation}
D = \lim_{u\rightarrow 0} \frac{u^2}{2d}\sum_m \langle ({\bf r}_m-{\bf r}_0)^2\tilde{P}({\bf r}_m,u|{\bf r}_0)
\rangle,
\label{eq:2}
\end{equation}
where
\begin{equation}
\tilde{P}({\bf r}_m,u|{\bf r}_0) = \int_0^{\infty} P({\bf r}_m,t|{\bf r}_0,0) e^{-ut} dt
\label{eq:3}
\end{equation}
is the Laplace transform of the transition probability $ P({\bf r}_m,t|{\bf r}_0,0)$, $d$ is the dimension of the space
and $\langle \cdots \rangle$ denotes an ensemble average over the random distribution of sites.
I exploit the coherent medium approximation for positionally disordered systems \cite{oda-lax, oda-lax-1d}.
I first divide the space around a site into $z$ equivalent cones and assume that a random walker makes a jump to
the adjacent site in one of the $z$ cones.
Within this approximation, the normalized diffusion constant is
given by
\begin{equation}
\frac{D}{D_0} =  \frac{w_c}{w_0},
\label{eq:4}
\end{equation}
where $D_0$ and $w_0$ are the diffusion constant and jump rate of
a reference regular system and
 $w_c$ is the coherent jump rate which is determined self-consistently by
\begin{equation}
\frac{2}{zw_c}= \int_0^{\infty}\frac{N(r) dr}{(\frac{z}{2} -1) w_c + w(r)}.
\label{eq:5}
\end{equation}
Here $N(r)$ represents the distribution function of the distance between
adjacent neighbors in a cone. In three dimensions,
\begin{equation}
N(r) = \frac{4\pi r^2 n}{z} \exp\left(-\frac{4\pi r^3 n}{3z}\right)
\label{eq:6}
\end{equation}
when sites are distributed randomly with density $n$.
In Eq.~(\ref{eq:4}), the scale of the length of the system under consideration is assumed to be the same as that of the reference system
since it does not play any significant role here.

%%Sect 3
\section{\label{Sec:3} Extended percolation in one dimension}
In one dimension, $z$ is set to $z=2$ in Eq.~(\ref{eq:5}) and
the self-consistency equation for the coherent jump rate $w_c$ reads as
\begin{equation}
\frac{1}{w_c}= \int_0^{\infty}\frac{N(x) dx}{ w(x)},
\label{eq:7}
\end{equation}
and the distribution function $N(x)$ becomes
\begin{equation}
N(x) = n e^{-nx}.
\label{eq:8}
\end{equation}
In this section, I investigate several different forms of $w(x)$
which is considered as a percolation model with long range connection,
and discuss possibility of a diffusive to non-diffusive transition.

\subsection{Simple percolation model}
As the simplest model, I first consider a transition probability
\begin{equation}
w(x) = \left\{ \begin{array}{ll}
w_0 & \mbox{(when $ 0 \le x \le {\bf r}_0$)}\\
\epsilon w_0 & \mbox{(when $ x > {\bf r}_0$ and $\epsilon \rightarrow 0$)} .
\end{array}
\right.
\label{eq:9}
\end{equation}
It is straightforward to obtain the diffusion constant from Eqs.~(\ref{eq:4})
and (\ref{eq:7}) $\sim$ (\ref{eq:9}).
I find
\begin{equation}
\frac{D}{D_0} = \frac{1}{1 + \frac{1-\epsilon}{\epsilon} e^{-n{\bf r}_0}},
\label{eq:10}
\end{equation}
and Fig.~1 shows the dependence of $D/D_0$ on the scaled density $nx_0$ for
 $\epsilon = 10^{-2}, 10^{-3}, 10^{-4}$.
As expected, the diffusion constant is identically zero in the percolation limit
$\epsilon = 0$.
When $\epsilon$ is finite, the diffusion constant is given by a sigmoid function whose
inflection point is at $nx_0 = -\ln[\epsilon/(1-\epsilon)]$.

\begin{figure*}
\begin{center}
\includegraphics[width=5cm]{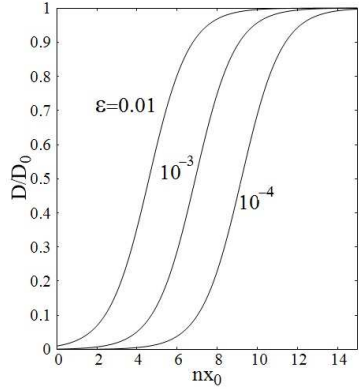}
%\vspace{4cm}
\caption{\label{fig1} The scaled diffusion constants of a simple percolation model
Eq.~(\ref{eq:9}) are shown as functions of the scaled density for
 $\epsilon = 10^{-2}, 10^{-3}, 10^{-4}$.}
\end{center}
\end{figure*}

\subsection{Extended percolation model}
I define an extended percolation model by a jump rate
\begin{equation}
w(x) = \left\{ \begin{array}{ll}
w_0 & \mbox{(when $ 0 \le x \le x_0$)}\\
w_0\left(\frac{x_0}{x} \right)^{\alpha} & \mbox{(when $ x > x_0$ )} ,
\end{array}
\right.
\label{eq:11}
\end{equation}
where $\alpha > 0$ is assumed. Namely, in this model, the range of jump beyond $x_0$
decays as a power-law function with exponent $-\alpha$.
From Eqs.~(\ref{eq:4}), (\ref{eq:7}), (\ref{eq:8}) and (\ref{eq:11}), I find
\begin{equation}
\frac{D}{D_0} =  \frac{1}{1-e^{-n{\bf r}_0} + (n{\bf r}_0)^{-\alpha} \Gamma(\alpha + 1, n{\bf r}_0) },
\label{eq:12}
\end{equation}
where $\Gamma(s,x) \equiv \int_x^{\infty}e^{-t} t^{s-1} dt$ is
the upper incomplete Gamma function.
Figure 2 shows the $nx_0$ dependence of the diffusion constant for $\alpha = 1, 5, 10, 20$.

The diffusion constant becomes identically zero at $\alpha = \infty$,
since $\alpha = \infty$ corresponds to the percolation limit.
\begin{figure*}[ht]
\begin{center}
\includegraphics[width=5cm]{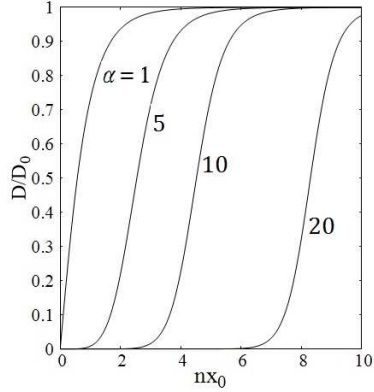}
%\vspace{4cm}
\caption{\label{fig2} Scaled diffusion constants for an extended percolation model
Eq.~(\ref{eq:11}) are shown as functions of the scaled density for
 $\alpha = 1, 5, 10, 20$.}
\end{center}
\end{figure*}

\subsection{Logistic-type model}
I consider a smooth function for the jump rate represented by a kind of the logistic curve
\begin{equation}
w(x) = w_0 \frac{e^{x_r/x_0}-1}{e^{x_r/x_0}+e^{x/x_0} - 2}.
\label{eq:13}
\end{equation}
which satisfies $w(0) = w_0$, $w(x_r) = w_0/2$ and $w(\infty) = 0$.
%%%as shown in Fig.~3 (a).
The diffusion constant is given by
\begin{equation}
\frac{D}{D_0} =1 - \frac{1}{1 + (e^{-x_r/x_0}-1)(nx_0 -1)},
\label{eq:14}
\end{equation}
Figure 3 shows the diffusion constant as functions of $nx_0$ for $x_r/x_0 =1.1, 2, 3$.
Consequently, there is a non-diffusive to diffusive transition at
$nx_0 = 1$.
Since $D/D_0 \simeq (e^{-x_r/x_0}-1)(nx_0 -1)$ when $nx_0 \sim 1$, the critical exponent of
the diffusion constant is unity.

\begin{figure*}[ht]
\begin{center}
%\includegraphics[width=5cm]{RWfig3a.eps}
%\hspace{1cm}
\includegraphics[width=5cm]{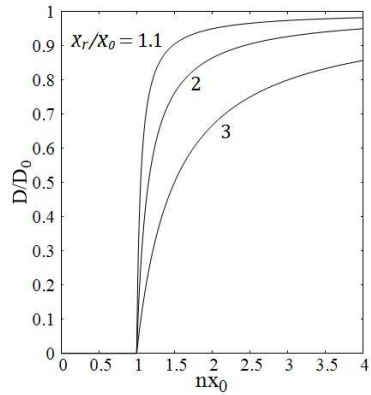}\\
%(a)\hspace{5cm}(b)
%\vspace{4cm}
\caption{\label{fig3} %(a) The jump rate of the logistic-type model. (b)
Scaled diffusion constants for a logistic-type model
Eq.~(\ref{eq:13}) are shown as functions of the scaled density for
 $x_r/x_0 =1.1, 2, 3$.
}
\end{center}
\end{figure*}

%%Sect 4
\section{\label{Sec:4} Extended percolation in three dimensions}

\subsection{Extended percolation model}
I consider an extended percolation model in three dimensions where sites are distributed randomly with density $n$ 
in a three dimensional space and
the jump rate $w(r)$ is given by
\begin{equation}
w(r) = \left\{ \begin{array}{ll}
w_0 & \mbox{(when $ 0 \le r \le r_0$)}\\
w_0\left(\frac{r_0}{r} \right)^{\alpha} & \mbox{(when $ r > r_0$ )}.
\end{array}
\right.
\label{eq:15}
\end{equation}

Self-consistency equations (\ref{eq:5}) and (\ref{eq:6}) for $w_c$ with Eq.~(\ref{eq:15}) are solved numerically.
Figure 4 shows the dependence of the scaled diffusion constant $D/D_0$ on the scaled
density $(4\pi r_0^3/3)n$ for $\alpha = 10$ and $\infty$, where $z=6$ is used as an example.
\begin{figure*}[h]
\begin{center}
\includegraphics[width=5cm]{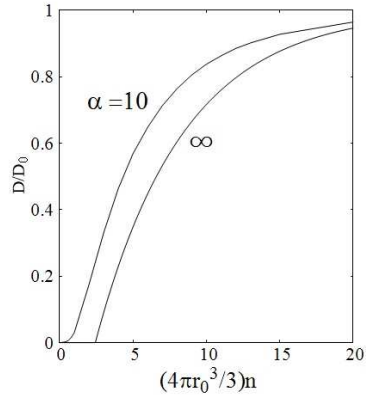}
%\vspace{4cm}
\caption{\label{fig4}The dependence of the scaled diffusion constant on the scaled density
for the extended percolation model for $\alpha = 10, \infty$.
Note $\alpha = \infty$ corresponds to the standard percolation model.
}
\end{center}
\end{figure*}
There are no percolation transition for $\alpha < \infty$.
The case $\alpha = \infty$ is the simple percolation model, the diffusion constant of which is given by
\begin{equation}
\frac{D}{D_0} = 1 - \frac{z}{z-2}\exp\left(- \frac{4\pi r_0^3}{3z} n\right)
\label{eq:16}
\end{equation}
and the critical percolation density is given by
\begin{equation}
\left(\frac{4\pi r_0^3}{3} n\right)_c = z \ln\frac{z}{z-2}
\label{eq:17}
\end{equation}
When $z=6$, $\left( \frac{4\pi r_0^3}{3} n\right)_c = 2.43$. 

\subsection{Super exponential decay model}
I consider the jump rate $w(r)$ given by
\begin{equation}
w(r) = \left\{ \begin{array}{ll}
w_0 & \mbox{(when $ 0 \le r \le r_0$)}\\
w_0\exp\{-[k(r-r_0)]^{\beta}\} & \mbox{(when $ r > r_0$ )}.
\end{array}
\right.
\label{eq:18}
\end{equation}
Figure 5 represents the dependence of the scaled diffusion constant $D/D_0$ on the scaled
density $(4\pi r_0^3/3)n$ for $\beta = 1, 4$, where $kr_0 =3$ and $z=6$ are used. 
\begin{figure*}[h]
\begin{center}
\includegraphics[width=5cm]{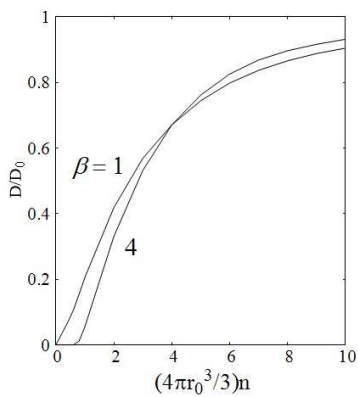}
%\vspace{4cm}
\caption{\label{fig5}The dependence of the scaled diffusion constant on the scaled density
for the super Gaussian decay model for $\beta = 1, 4$.}
\end{center}
\end{figure*}
Apparently, there is a non-diffusive to diffusive transition for $\beta = 4$.
In fact, a non-diffusive to diffusive transition exists when $\beta \ge 3$.

%%Sect 5
\section{\label{Sec:5}Discussion}
I have studied random walks where a random walker can make long range jumps
and obtained characteristic behavior of the diffusion constant
within the coherent medium approximation.
As for the distance dependence of the jump rate, I investigated
different types of extended percolation models.
It is shown that a non-diffusive to diffusive transition exists
in certain types of the jump rate function in one and three dimensions.

The self-consistency Eq.~(\ref{eq:5}) supports a solution $w_c = 0$ only when
\begin{equation}
\int_0^{\infty}\frac{N(r) dr}{w(r)} = \infty.
\label{eq:19}
\end{equation}
Therefore, in one dimension, the non-diffusive to diffusive transition
exists when the jump rate function exhibits exponential or faster decay
in the long distance limit \cite{oda-lax-1d}.
In $d$-dimensional systems. there are no non-diffusive states
unless the jump rate function decays faster than $e^{-r^d}$. 

The present results will give some insights in random walks
in a highly complex structure like the free energy landscape.

%\vspace{0.5cm}
%\noindent
{\large\bf Acknowledgments}\\
I would like to thank Dr. M. F. Shlesinger for valuable discussion.
This work was supported in part by JSPS KAKENHI Grant Number 
18K03573.

\end{document}